\pdfoutput=1
\documentclass[modern]{aastex63_arxiv}

\usepackage{algorithm}
\usepackage{algpseudocode}
\usepackage{amsmath}
\usepackage{booktabs}
\usepackage{hyperref}
\usepackage{glossaries}
\usepackage{graphicx}
\usepackage{longtable}
\usepackage{xcolor}

\usepackage{glossaries} 
\setacronymstyle{long-short} 
\newacronym{cmd}{CMD}{color-magnitude diagram}
\newacronym{mcmc}{MCMC}{Markov chains Monte Carlo}
\newacronym{rcg}{RCG}{red clump giant}

\usepackage{amssymb}    

\DeclareUnicodeCharacter{0308}{}  

\newcommand{\asnow}{a_\mathrm{snow}}
\newcommand{\au}{\mathrm{au}}

\newcommand{\days}{\mathrm{d}}

\newcommand{\DL}{D_\mathrm{L}}
\newcommand{\DS}{D_\mathrm{S}}
\newcommand{\eg}{e.g.}

\newcommand{\Eq}[1]{Equation~\ref{#1}}
\newcommand{\eventname}{MOA-2014-BLG-472}

\newcommand{\eventnameLb}{MOA-2014-BLG-472Lb}
\newcommand{\Fig}[1]{Figure~\ref{#1}}
\newcommand{\fbl}[1]{f_{\mathrm{b},#1}} 
\newcommand{\fsl}[1]{f_{\mathrm{s},#1}} 
\newcommand{\G}{G}
\newcommand{\g}[1]{g_\mathrm{#1}}
\newcommand{\ie}{i.e.}
\newcommand{\Is}{I_\mathrm{S}}
\newcommand{\Iso}{I_\mathrm{S,0}}

\newcommand{\Irc}{I_\mathrm{RCG}}
\newcommand{\kpc}{\mathrm{kpc}}

\newcommand{\mas}{\mathrm{mas}}
\newcommand{\meter}{\mathrm{m}}

\newcommand{\Mjup}{\mathrm{M}_\mathrm{J}}
\newcommand{\Msun}{\mathrm{M}_\odot}

\newcommand{\ML}{M}
\newcommand{\MLa}{M_1}
\newcommand{\MLb}{M_2}

\newcommand{\murelg}{\mu_\textrm{rel,G}}
\newcommand{\p}{\,.}
\newcommand{\piE}{\pi_\mathrm{E}}
\newcommand{\piEv}{\boldsymbol{\pi}_\mathrm{E}}
\newcommand{\piEE}{\pi_{\mathrm{E},E}}
\newcommand{\piEN}{\pi_{\mathrm{E},N}}

\newcommand{\pirel}{\pi_\mathrm{rel}}

\newcommand{\RMOA}{R_\mathrm{MOA}}
\newcommand{\Rsun}{\mathrm{R}_\odot}

\newcommand{\Sec}[1]{Section~\ref{#1}}
\newcommand{\Tab}[1]{Table~\ref{#1}}

\newcommand{\thE}{\theta_\mathrm{E}}
\newcommand{\thS}{\theta_\star}
\newcommand{\tE}{t_\mathrm{E}}

\newcommand{\tS}{t_\star}

\newcommand{\tento}[1]{10^{#1}}
\newcommand{\thjd}{\mathrm{HJD^\prime}}
\newcommand{\uas}{\mu\mathrm{as}}
\renewcommand{\v}{\,,}

\newcommand{\vmi}[1]{\left(V-I\right)_\mathrm{#1}}
\newcommand{\vmirc}{(V-I)_\mathrm{RCG}}

\newcommand{\vpm}[3]{#1^{+#2}_{-#3}}
\newcommand{\yr}{\mathrm{yr}}

\newcommand{\murel}{\mu_\mathrm{rel}}

\newcommand\ltsima{$\; \buildrel <\over\sim \;$}
\newcommand\simlt{\lower.5ex\hbox{\ltsima}}
\newcommand\gtsima{$\; \buildrel >\over\sim \;$}
\newcommand\simgt{\lower.5ex\hbox{\gtsima}}

\renewcommand{\phn}{\phantom{0}}
\newcommand{\phntw}{\phn\phn}
\newcommand{\phnth}{\phn\phn\phn}
\newcommand{\phnfo}{\phn\phn\phn\phn}

\usepackage[T1]{fontenc}
\usepackage{ae,aecompl}
\usepackage{newtxtext,newtxmath}

\newcommand{\affa}{Sorbonne Université, CNRS, UMR 7095, Institut d'Astrophysique de Paris, 98 bis bd Arago, 75014 Paris, France}
\newcommand{\affb}{Code 667, NASA Goddard Space Flight Center, Greenbelt, MD 20771, USA}
\newcommand{\affc}{Department of Astronomy, University of Maryland, College Park, MD 20742, USA}
\newcommand{\affd}{Department of Astronomy, Graduate School of Science, The University of Tokyo, 7-3-1 Hongo, Bunkyo-ku, Tokyo 113-0033, Japan}
\newcommand{\affe}{National Astronomical Observatory of Japan, 2-21-1 Osawa, Mitaka, Tokyo 181-8588, Japan}
\newcommand{\afff}{Astronomical Observatory, University of Warsaw, Al. Ujazdowskie 4, 00-478 Warszawa, Poland}
\newcommand{\affg}{Department of Earth and Space Science, Graduate School of Science, Osaka University, Toyonaka, Osaka 560-0043, Japan}
\newcommand{\affh}{Institute of Natural and Mathematical Sciences, Massey University, Auckland 0745, New Zealand}
\newcommand{\affi}{Laboratory for Particle Physics and Cosmology, Harvard University, Cambridge, MA, United States of America}
\newcommand{\affj}{Department of Applied Mathematics and Theoretical Physics, University of Cambridge, Wilberforce Road, Cambridge, CB3 0WA, United Kingdom}
\newcommand{\affk}{Institute for Space-Earth Environmental Research, Nagoya University, Nagoya 464-8601, Japan}
\newcommand{\affl}{Department of Physics, University of Auckland, Private Bag 92019, Auckland, New Zealand}
\newcommand{\affm}{Department of Earth and Planetary Science, Graduate School of Science, The University of Tokyo, 7-3-1 Hongo, Bunkyo-ku, Tokyo 113-0033, Japan}
\newcommand{\affn}{Instituto de Astrof\'isica de Canarias, V\'ia L\'actea s/n, E-38205 La Laguna, Tenerife, Spain}
\newcommand{\affo}{Department of Physics, The Catholic University of America, Washington, DC 20064, USA}
\newcommand{\affp}{Code 660, NASA Goddard Space Flight Center, Greenbelt, MD 20771, USA}
\newcommand{\affq}{Universities Space Research Association, Columbia, MD 21046, USA}
\newcommand{\affr}{University of Canterbury Mt.\ John Observatory, P.O. Box 56, Lake Tekapo 8770, New Zealand}
\newcommand{\affs}{Department of Physics, Faculty of Science, Kyoto Sangyo University, 603-8555 Kyoto, Japan}

\received{2021 March 12}
\revised{2021 June 11}
\accepted{Accepted 2021 June 17}
\published{2021 June 26}

\begin{document}

\title{\bfseries New Giant Planet beyond the Snow Line for an \\Extended MOA Exoplanet Microlens Sample}

\correspondingauthor{Cl{\'e}ment Ranc}
\email{ranc@iap.fr}

\author[0000-0003-2388-4534]{Cl{\'e}ment Ranc}
\affiliation{\affa}

\author[0000-0001-8043-8413]{David~P.~Bennett}
\affiliation{\affb}
\affiliation{\affc}
\author[0000-0003-4916-0892]{Richard~K.~Barry}
\affiliation{\affb}
\author[0000-0003-2302-9562]{Naoki~Koshimoto}
\affiliation{\affd}
\affiliation{\affe}
\author[0000-0002-2335-1730]{Jan~Skowron}
\affiliation{\afff}
\author{Yuki~Hirao}
\affiliation{\affg}
\author{Ian~A.~Bond}
\affiliation{\affh}
\author[0000-0002-4035-5012]{Takahiro~Sumi}
\affiliation{\affg}
\author[0000-0002-4861-6192]{Lars~Bathe-Peters}
\affiliation{\affi}
\affiliation{\affj}
\affiliation{\affg}
\author{Fumio~Abe}
\affiliation{\affk}
\author{Aparna~Bhattacharya}
\affiliation{\affb}
\affiliation{\affc}
\author{Martin~Donachie}
\affiliation{\affl}
\author{Hirosane~Fujii}
\affiliation{\affg}
\author[0000-0002-4909-5763]{Akihiko~Fukui}
\affiliation{\affm}
\affiliation{\affn}
\author[0000-0003-2267-1246]{Stela Ishitani Silva}
\affiliation{\affo}
\affiliation{\affb}
\author[0000-0002-8198-1968]{Yoshitaka~Itow}
\affiliation{\affk}
\author{Rintaro~Kirikawa}
\affiliation{\affg}
\author[0000-0002-3401-1029]{Iona~Kondo}
\affiliation{\affg}
\author{Man~Cheung~Alex~Li}
\affiliation{\affl}
\author{Yutaka~Matsubara}
\affiliation{\affk}
\author[0000-0003-1978-2092]{Yasushi~Muraki}
\affiliation{\affk}
\author[0000-0001-9818-1513]{Shota~Miyazaki}
\affiliation{\affg}
\author[0000-0001-8472-2219]{Greg Olmschenk}
\affiliation{\affp}
\affiliation{\affq}
\author[0000-0001-5069-319X]{Nicholas~J.~Rattenbury}
\affiliation{\affl}
\author{Yuki~Satoh}
\affiliation{\affg}
\author{Hikaru~Shoji}
\affiliation{\affg}
\author[0000-0002-5843-9433]{Daisuke~Suzuki}
\affiliation{\affg}
\author{Yuzuru~Tanaka}
\affiliation{\affg}
\author{Paul~J.~Tristram}
\affiliation{\affr}
\author{Tsubasa~Yamawaki}
\affiliation{\affg}
\author[0000-0003-3480-0973]{Atsunori~Yonehara}
\affiliation{\affs}

\begin{abstract}
Characterizing a planet detected by microlensing is hard if the planetary signal is weak or the lens-source relative trajectory is far from caustics.
However, statistical analyses of planet demography must include those planets to accurately determine occurrence rates. As part of a systematic modeling effort in the context of a $>10$-year retrospective analysis of MOA’s survey observations to build an extended MOA statistical sample, we analyze the light curve of the planetary microlensing event MOA-2014-BLG-472. This event provides weak constraints on the physical parameters of the lens, as a result of a planetary anomaly occurring at low magnification in the light curve. We use a Bayesian analysis to estimate the properties of the planet, based on a refined Galactic model and the assumption that all Milky Way's stars have an equal planet-hosting probability. We find that a lens consisting of a $\vpm{1.9}{2.2}{1.2}\,\Mjup$ giant planet orbiting a $\vpm{0.31}{0.36}{0.19}\,\Msun$ host at a projected separation of $0.75\pm0.24\,\au$ is consistent with the observations and is most likely, based on the Galactic priors. The lens most probably lies in the Galactic bulge, at $\vpm{7.2}{0.6}{1.7}\kpc$ from Earth. The accurate measurement of the measured planet-to-host star mass ratio will be included in the next statistical analysis of cold planet demography detected by microlensing.
\end{abstract}


\section{Introduction} \label{sec:intro}

During fall 2020, the hundredth exoplanet detection through gravitational microlensing
was added to the NASA Exoplanet Archive database\footnote{https://exoplanetarchive.ipac.caltech.edu/}. Although modest in amount
when compared to the 4,379 confirmed exoplanets to date and distributed in more
than 3,237 planetary systems \citep{Schneider.2011}, this milestone enables an
unprecedented look at the demography of cold exoplanets orbiting their host stars on
wide orbits,  with a typical semi-major axis of $\sim 0.5\text{-}10\,\au$.
Microlensing detections dominate the population of confirmed planets 
below one Saturn mass and located beyond the ``snow line'', \ie, the inner boundary of
the protoplanetary disk where planet formation is most efficient, according to 
the core accretion theory \citep{Lissauer.1987, Lissauer.1993, Pollack.1996}.
So, this sample represents a relatively new and unique opportunity for planet
formation theories to compare predictions with observations, in a region of 
the parameter space largely unexplored by other planet detection techniques.

The first comparison of the microlensing planet occurrence rate with population
synthesis models \citep{Ida.2004,Mordasini.2018} identified a discrepancy
between predictions of the core accretion theory's runaway gas accretion
process and observations \citep{Suzuki.2018}. In particular, the observational
results do not show any dearth of intermediate-mass giant planets, while the
models predict 10 times fewer planets  in the planet-to-host mass ratio range
$10^{-4}<q<4\times10^{-4}$. Resolving this discrepancy may have important
implications in our understanding of the role played by the runaway gas
accretion phase in the delivery of water to inner planetary orbits
\citep{Raymond.2017}. The MOA collaboration is currently performing a
systematic retrospective analysis including more than ten years of survey observations performed
at the Mount John in New Zealand, to strengthen and expand the previous
statistical results on microlensing planet occurrence rate \citep{Gould.2010,
    Sumi.2010, Cassan.2012, Shvartzvald.2016, Suzuki.2016, Udalski.2018}.

So far, this systematic analysis of previous survey data led to the discovery
of several missed exoplanets \citep[\eg,][]{Kondo.2019}. The discovery
presented in this article takes place in the context of this systematic
modeling of past detections. We report the discovery of a new giant planet from
the analysis of the microlensing event MOA-2014-BLG-472, initially detected by
alert systems. The planetary signal for this event is not created by a caustic
crossing. As a result, the planetary anomaly in the light curve has a low
magnification, and the constraints on the physical parameters of the lens are
weak. 
However, including planets like MOA-2014-BLG-472Lb in statistical studies on
planet demography is crucial for the completeness of planetary occurrence rates.

This article describes the full analysis of \eventname. In Section~\ref{sec:obs}, we
recount the discovery of the event, describe the observations and select the
data set used to model the event. In Section~\ref{sec:models}, we describe the
full light-curve modeling process. In Section~\ref{sec:properties}, we use a
galactic model to derive the physical properties of the source and lens.
Section~\ref{sec:conclusion} provides a summary of the analysis and concludes
the article.

\section{Observations and data reduction}\label{sec:obs}

\begin{figure}[tp]
  \begin{center}
    \includegraphics[scale=1.3]{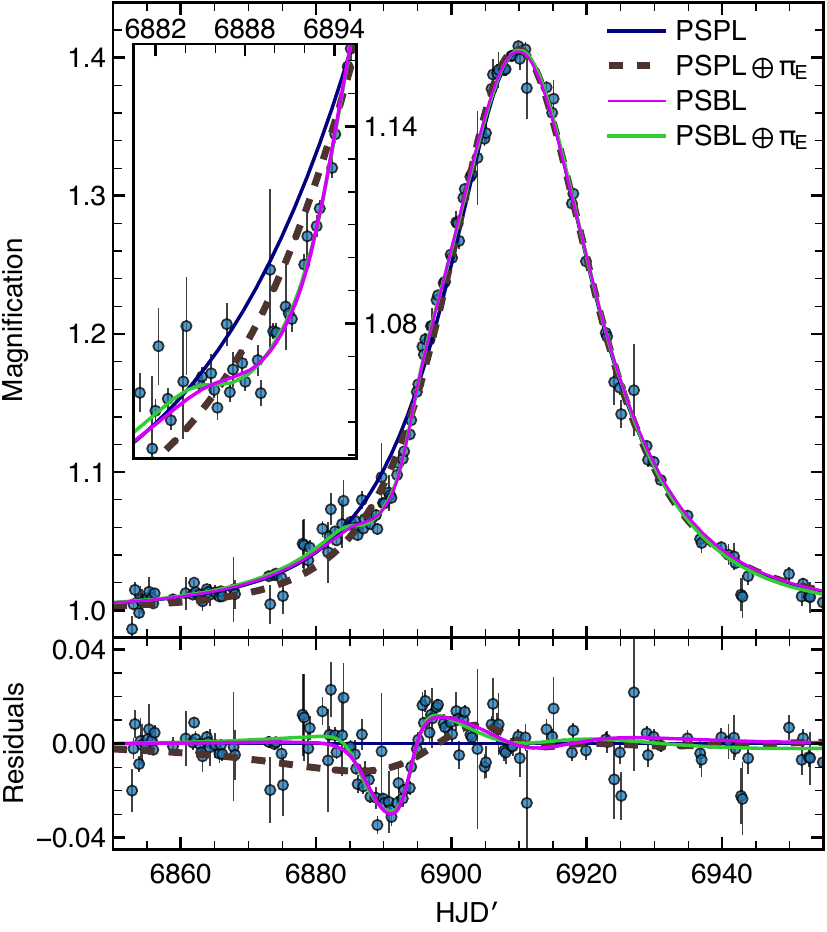}
    \caption{Magnification during the microlensing event MOA-2014-BLG-472 and the best point-source binary-lens model (PSBL; solid magenta line). For comparison, the black solid line, brown dashed line and green solid line respectively show the rejected point-source point-lens (PSPL), PSPL including parallax, and PSBL with parallax models. MOA observations are displayed in blue. The inset shows a zoom in on the anomaly. In the lower panel, the residuals from the PSPL model are plotted in magnification units. For clarity, the plot uses 5-hour bins to display the data (these bins are not used in the fit).}\label{fig:lc1}
  \end{center}
\end{figure}

The microlensing event \eventname{} was discovered by the Microlensing
Observations in Astrophysics \cite[MOA, phase II;][]{Sumi.2003} collaboration
and first alerted on 2014 August 16 at UT~11:40, \ie,
$\thjd\approx6885.99$\footnote{$\thjd=\mathrm{HJD}-2,450,000$.}. The event is
located at the J2000 equatorial coordinates $\left(\mathrm{RA},\
\mathrm{Dec.}\right)=(18^\mathrm{h}\, 00^\mathrm{m}\, 19\overset{^\mathrm{s}}{.}40,\ -28\arcdeg\, 08\arcmin\,
56\overset{\prime\prime}{.}29)$ in the MOA-II field `gb10.' MOA observations were performed
using a $1.8\,\meter$ telescope \citep[and its $2.2\,\mathrm{deg}^2$ field of
view camera,][]{Sako.2008} at the Mount John University Observatory in New
Zealand with a cadence of 15\,min in the custom wide-band MOA $R/I$-band
filter, referred to as $\RMOA$. An anomaly was detected in real time by the 
MOA observers who issued an internal MOA alert on 2014 September 4. 
MOA's implementation of the DIA method \citep{Bond.2001} has been used to
extract the photometry of MOA observations.

The Optical Gravitational Lensing Experiment \citep[OGLE, phase IV;][]{Udalski.2015} was also monitoring this event and triggered an alert on the Early Warning System (EWS) website on 2014 August 26 at UT~11:06, naming the event OGLE-2014-BLG-1783. This event lies in the OGLE-IV field `BLG504.08,' and has been observed with the $1.3\,\meter$-telescope located at Las Campanas Observatory in Chile (and its $1.4\,\mathrm{deg^2}$ field of view camera), with a cadence of $1\,\mathrm{hour}^{-1}$.
The anomaly has been detected by OGLE independently in their data, and an internal alert was sent on 2014 August 26 at UT~11:22. OGLE's member Jan Skowron circulated among all the collaborations the first model performed, in real time, indicating a likely planet (with mass ratio of 0.0056) on 2014 September 20.

The final data sets consist of $13,789$ data points from MOA observations and used to model the microlensing light curve. We select five observing seasons (2 before and 2 after the event's year) to prevent missing some potential 
variability in the baseline.
The microlensing event has a weak maximal amplification of only $0.25\,\mathrm{mag}$. However, due to the source star being a Red Clump giant ($I\sim15.2\,\mathrm{mag}$), it is still well detectable/observable.
\Fig{fig:lc1}, shows the magnification of the source flux as a function of time. The peak of magnification occurs at $\thjd\approx6910$, and a clear anomaly starts at $\thjd\approx6885$, first slowing down the magnification rise, then suddenly hiking up the magnification faster than a single-lens magnification pattern.
Moreover, the anomaly
occurs at an extremely low magnification, $A<1.1$. The \Fig{fig:lc1} displays 5-hour bins for clarity purposes, but all the data are used during the modeling process.

As a consequence, the error bars are
expected to play a major role in the final uncertainties on the physical parameters.
Since the photometry pipelines typically underestimate the error bars,
for each data set, we normalized the error bars on magnitudes, $\sigma$, so that the $\chi^2$ per degree of freedom
is $\chi^2/\mathrm{d.o.f.}=1$, and the cumulative sum of $\chi^2$ is approximately linear. This
procedure assumes a best-fit model, and can be repeated as new plausible models are found.
During the broad initial search in the parameter space, the error bars are not changed. Then,
while exploring local $\chi^2$-minima, we use
the normalization law \citep{Yee.2012}
\begin{equation}
    \sigma^\prime_i = k\sqrt{\sigma^2+e_\mathrm{min}^2},
\end{equation}
where $\sigma^\prime$ is the normalized error bar, the constant $k$ is the rescaling factor, and the constant $e_\mathrm{min}$ mostly modifies the highly magnified data. For \eventname, we use $k=1.205$ and $e_\mathrm{min}=2.763\times10^{-3}$.

\section{Light-curve Models}\label{sec:models}

\subsection{Single-lens Model}\label{sec:single}

We start modeling the event \eventname{} by fitting the observations with a
Paczy\'nski light curve \citep{Paczynski.1986}, described by three independent
parameters: the time ($t_0$) and projected separation ($u_0$) when lens and
source are closest on the sky, and the Einstein radius crossing time,
\begin{equation}\label{eq:tE}
    \tE = \frac{\thE}{\murel}\v
\end{equation}
where $\murel$ is the lens-source relative proper motion in the geocentric
reference frame and $\thE$ is the angular Einstein radius.
These three parameters can be approximately estimated without
any sophisticated numerical techniques.
First, the peak of the event shown in \Fig{fig:lc1} provides $\thjd \approx 6910$.
Second, the peak-to-baseline flux
ratio provides an estimate of the magnification at the peak of the event,
$A_\mathrm{peak}\approx 1.3$.
Using Taylor series for the expression of the magnification
yields $u_0\approx 1$ at the peak.
Third, we derive the expected magnification at $t=t_0+\tE$, and search for the
corresponding flux in the light curve to find $\tE\approx12\,\days$.

Above, we derived estimates for model parameters by assuming that the
flux measurement comes entirely from the source star, which is almost never true.
During the modeling process, the three parameters $t_0$, $u_0$ and $\tE$ 
yield the magnification at any given date. To evaluate the goodness-of-fit of
a model, two additional parameters are required to compute the observable: one describes 
the unlensed source flux
$\fsl{\lambda_i}$, for any passband, $\lambda_i$; the other is the excess flux,
$\fbl{\lambda_i}$, resulting from the combination of any (and possibly
several) `blend' stars.
The blend can be either the lens itself or an unrelated star or stars.
At any time $t$, the total flux of the
microlensing target is
\begin{equation}\label{eq:ftotal}
  \Phi_{\lambda_i} (t) = A(t) \fsl{\lambda_i} + \fbl{\lambda_i}\v
\end{equation}
where $A(t)$ is the source flux magnification at the date $t$, and
$\lambda_i$ is the MOA $R$ passband.

Starting from the parameter estimated above, we use a Levenberg–Marquardt
algorithm \citep{Levenberg.1944} to find the best fit model parameters to be
used as a starting position when searching for binary-lens models. We then use
a \gls*{mcmc} algorithm to determine the uncertainties.
At this stage, we remove the data during the anomaly,
    since a point-source point-lens model (hereafter `PSPL') cannot (by definition) produce any anomaly.
The median parameters and their credible intervals are: %
$t_0 = 6910.1 \pm 0.1$,
$u_0 = 0.9 \pm 0.1$, and
$\tE = 14 \pm 1\,\days$. 
For comparison with the other models 
presented in this article, the $\chi^2$ value computed with the entire dataset is $\chi^2=14744$.
The best fit PSPL
model is shown with a solid dark blue line in \Fig{fig:lc1}.
In this figure, the data are
binned for more clarity. We choose 5-hour bins, such that 
for each bin, $n$, consisting of $\mathrm{N}_n$ data, the plotted uncertainty, $\sigma_n^{\prime\prime}$, and
magnification, $A_n$, are defined as
\begin{equation}
    \sigma_n^{\prime\prime}=\left(\sum_{j=1}^{\mathrm{N}_n} \sigma_j^{\prime\, 2}\right)^{-1/2}\quad \text{and}\quad
    A_n = \sigma_n^{\prime\prime\,2} \sum_{j=1}^{\mathrm{N}_n}\frac{A_j}{\sigma_j^{\prime\, 2}}\p
\end{equation}
We do not use any binned data during the fitting process, though.

We introduce two additional parameters to assess whether the anomaly in the light curve may be
explained by the non-inertial nature of the observer reference frame. These are the Northern and Eastern components of the microlens parallax vector in the geocentric frame, $\piEv$, respectively $\piEN$ and $\piEE$, as defined in \cite{Gould.2004}. The direction of vector $\piEv$ is the same as the instantaneous lens-source relative proper motion at $\thjd=6910$, and its magnitude is the lens-source relative parallax in units of the angular Einstein ring radius, \ie,
\begin{equation}\label{eq:piE}
    \piE = \frac{\pirel}{\thE}\v
\end{equation}
where $\pirel = 1\,\au/\DL - 1\,\au / \DS$, $\DL$ is the distance to the lens and $\DS$ the distance to the source. Starting from the best fitting static model, we use a \gls*{mcmc} to find the best model with parallax, and estimate the uncertainties. We now include all the observations, since we search for a parallax signal that could explain the anomaly.
Including the parallax in the model improves the $\chi^2$ by $\Delta\chi^2 = -380$. The median and credible intervals of the parameters are: 
$t_0 = 6909.80 \pm 0.05$,
$u_0 = 2.8 \pm 0.4$,
$\tE = \vpm{6.2}{0.8}{0.7}\,\days$,
$\piEN = 0 \pm 2$, and
$\piEE = 2.2 \pm 0.3$.
The results from the \gls*{mcmc} show that the constraint on $\piEN$ is very weak, allowing a broad range of acceptable values, including the solution $\piEN=0.0$ at a level $<1$-$\sigma$. The very large value of $\piE = \vpm{2.7}{1}{0.5}$ results from the inability for the model to fit the anomaly. This can be seen in \Fig{fig:lc1} that shows the best fit PSPL
model with microlens parallax (hereafter `$\text{PSPL}\oplus\piE$') with a thick brown dashed line. The static binary-lens model presented in \Sec{sec:binary-models} is preferred by $\Delta\chi^2=-588$ and fits the anomaly.

\begin{figure}[tp]
  \begin{center}
    \includegraphics[scale=1]{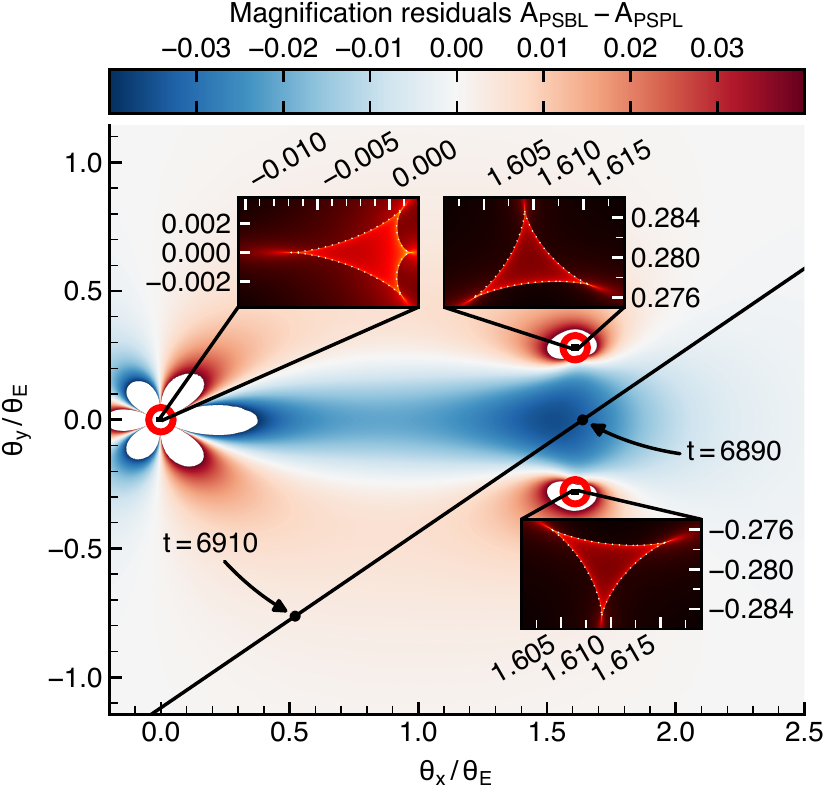}
    \caption{Caustic topology of the best-fit PSBL model. The color scale of the main plot refers to the magnification difference between the best PSBL and best PSPL models (except in the white regions around the caustic components, where the residuals are highest and not displayed). The caustic consists of three parts, located in the red circles. The insets display a zoom in on magnification maps in the vicinity of the caustic components in the source plane (dotted line). For convenience, we use two different logarithmic color scales for the central and planetary caustics. The black line shows the source-lens trajectory, and the black dots the source position at $\thjd=6890$ (anomaly peak) and $\thjd=6910$ (event peak). Blue regions denote a de-magnification compared to a single lens.}\label{fig:caus1}
  \end{center}
\end{figure}

\begin{figure}[tp]
  \begin{center}
    \includegraphics[width=\linewidth]{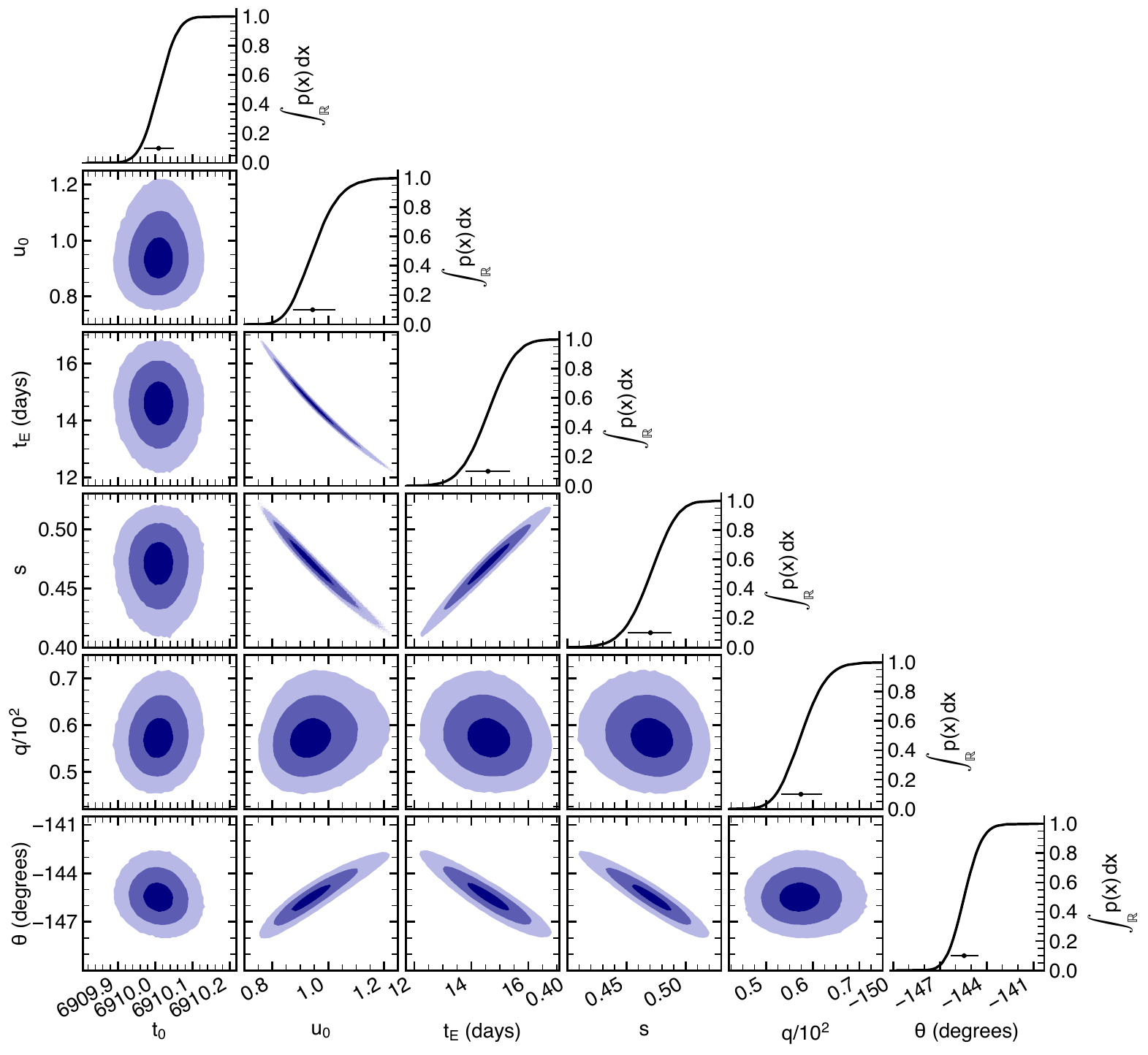}
    \caption{Correlation between the parameters for the PSBL model. The three shaded areas show the 1-3$\sigma$ credible regions, respectively, from the darkest to lightest color. Plots in the diagonal display the marginal cumulative distribution of each parameter (solid line), the median of the distribution (dot), and the 68.3 per cent credible interval centered on the median. Plot prepared using the python package \texttt{MOAna} \citep{Ranc2020moana}.}\label{fig:corr1}
  \end{center}
\end{figure}

\subsection{Binary-lens Models}\label{sec:binary-models}

\begin{table}[htp]
{\centering
\caption{Parameters for the Best-fit PSBL Model and the
    Corresponding Statistical Values from the Posterior Probability
    Distribution Function \label{tab:model_parameters}}\label{tab:observ}
{\footnotesize
\begin{tabular}{@{} l @{\hspace*{6pt}}l @{\hspace*{8pt}} c c @{}}\toprule
    Parameter              & Units   & Best-fit & MCMC results$^{(1)}$ \\ \midrule
$\chi^2$     &         & 13776.36\phnfo    & \dots                     \\ 
$q/10^{2}$   &         & \phnfo0.568450    & $0.575_{-0.042}^{+0.045}$ \\ 
$s$          &         & \phnfo0.475423    & $0.47\pm0.02$             \\ 
$\tE$        & days    & \phnth14.787693   & $14.6\pm0.8$              \\ 
$t_0$        & $\thjd$ & \phn6910.008646   & $6910.01\pm0.04$          \\ 
$u_0$        &         & \phnfo0.924277    & $0.94_{-0.07}^{+0.08}$    \\ 
$\alpha$     & deg     & \phn$-145.685730$ & $-145.4\pm0.9$            \\
$R_\mathrm{MOA,s}\,^{(2)}$ & & \phntw$-11.846185$  &  $-11.9\pm0.2$  \\
$\fbl{R}/\fsl{R}\,^{(3)}$ & & \phnfo$0.454891$  &  $0.4\pm0.2$  \\\midrule
$\Is$        &         & \dots  &  $15.8\pm0.2$  \\ \bottomrule \addlinespace[6pt]
\end{tabular}}\par}
{\footnotesize$^{(1)}$ Median of the marginalized posterior distributions, with error bars displaying the 68.3 per cent credible interval around the median.\\
$^{(2)}$ Instrumental source magnitude in MOA $R$-band filter.\\$^{(3)}$ Ratio of MOA $R$-band instrumental blend and source flux. We do not convert the blend flux from the $R$ to the $I$ passband because the nature of the blend is unknown.}
\end{table}

In \Sec{sec:single}, we showed that the event \eventname{} is not well described by a single-lens model, because the anomaly that occurs at $t\approx6890$ cannot result from a parallax effect on a single lens. Hence, we search for plausible binary-lens,
single source models. Three additional parameters are required: the mass ratio
of the secondary to primary lens component $q = \MLb/\MLa$, where $\MLb$
($\MLa$) is the mass of the secondary lens (the mass of the primary lens,
respectively); the separation in Einstein radius, $s$; and the counterclockwise
angle of the lens-source relative motion projected onto the sky plane with the
lens binary axis (from the secondary to the primary lens), $\alpha$. For a
binary lens model, we choose $u_0$ as the distance of closest approach between
the lens center of mass and the source.

\begin{figure}[tp]
  \begin{center}
    \includegraphics[scale=1]{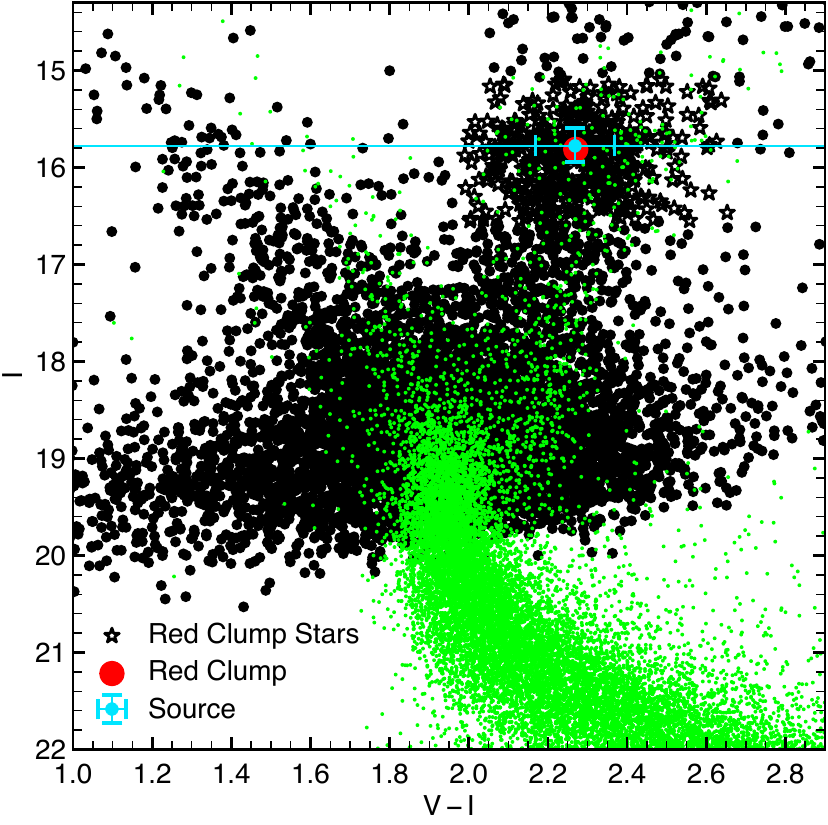}
    \caption{The black dots show a CMD in the standard Kron-Cousins $I$ and Johnson $V$ photometric systems of OGLE-III stars selected within 2~arcmin around the target. The RCG centroid is indicated by the red circle, and the RC stars are shown as black stars. The source brightness and the estimated source color are shown by the cyan point. The color dispersion of the RC stars mostly accounts for the source color error. The cyan horizontal line reminds us that we only measured the source brightness, and the source color fully follows from the assumption that the source belongs to the RC. The green dots show the Hubble Space Telescope CMD from \citet{Holtzman.1998} shifted to the bulge distance and extinction derived for the MOA-2014-BLG-472 line of sight.}\label{fig:cmd}
  \end{center}
\end{figure}

We start exploring the parameter space searching for PSBL solutions using the
best-fit PSPL model parameters, and the initial condition grid search method
introduced in \cite{Bennett.2010}. In practice, for each set of $\{s, q\}$, we scan over $-\pi \leq \alpha \leq \pi$
with a $1.1^\circ$ step. During this process, the best-fit PSPL model parameters
$\{t_0,u_0,\tE\}$ are kept fixed. We used 8 grid points in $\log{\epsilon}$, from 
$-4.0$ to $-0.5$, with a 0.5 grid spacing, where $\epsilon=q / (1+q)$ is the planetary mass fraction.
The separation values range from 0.1 to 10.0, evenly spaced on a grid of $\log{s}$, that includes:
\begin{itemize}
    \item 53 grid points  for $\log{\epsilon} \leq -2.0$,
    \item 70 grid points  for $\log{\epsilon} = -1.5$,
    \item 85 grid points  for $-1.0 \leq \log{\epsilon}$.
\end{itemize}
For each model, we compute the $\chi^2$ value and start 25 new fits from the 
best 25 models found on the grid. We only select one initial condition 
per $\{s, q\}$ couple, \ie, we use the best $\alpha$ value for a given set of $\{s, q\}$.
At this stage, we release all the parameter constraints, and we use an
adaptative version of the Metropolis algorithm optimizing the size of the
proposal function during the exploration of the parameter space with a Monte
Carlo method. The analysis of this set of fits leads us to identify four different models that meet our criterion
$\Delta\chi^2=\chi^2-\chi^2_\mathrm{min}\leq500$, for further in depth investigation.
We use these models to define four classes of models in the next step, consisting in
sampling the posterior distribution using several \gls*{mcmc} chains.

The two best fitting models have the same caustic topology, with close values of
$s$ and $q$. One is the best-fit model presented in \Tab{tab:model_parameters}. According to
the second class of best models ($\Delta\chi^2\approx 115$), the magnification peak would be due to one off-axis planetary caustic
characterized by $s=0.62798$ and $q=8.8766\cdot\tento{-3}$. However, this
model does not fit the anomaly: $\sim80\%$ of the $\chi^2$ difference compared to the
best-fit model comes from observations during the anomaly, and $\sim 20\%$ comes
from data between the anomaly and the event peak. This particular model is simply unable to 
reproduce the gradient of magnification during the anomaly, and must be rejected.
The third class of models ($\Delta\chi^2\approx 153$) involves a wide separation caustic. In this scenario,
the main peak of the event is due to the central caustic, the source trajectory is passing in between the two components of the caustic, but this model does not properly fit the gradient of magnification
during the anomaly: $\sim69\%$ of the $\chi^2$ increase compared to the best-fit model
occurs during the anomaly, and $24\%$ between the anomaly and the peak of the event. It is worth
noting that the description of the tails of the event given by this model is also poorer.
The fourth best model is substantially worse than the three others,
does not fit the anomaly, nor the event peak, and is characterized by $\Delta\chi^2\approx423$
(98.6\% of this value comes from data in the interval $\thjd=6865$-6910).

After checking the convergence of the
\gls*{mcmc} chains, we use $50,000$ samples to diagonalize the covariance matrix and
optimize the posterior sampling. \Fig{fig:caus1} displays the source trajectory
relative to the caustics obtained with the best-fit model. \Tab{tab:model_parameters} shows the median of
the marginalized posterior distributions. The error bars correspond to the
68.3 per cent credible interval around the median, derived from the 16 and 84 per cent
percentile of the one dimensional marginalized posterior distribution.
One-dimensional cumulative functions and two-dimensional covariances (and
nonlinearities) between the model parameters are shown in \Fig{fig:corr1}.

\Tab{tab:model_parameters} and \Fig{fig:corr1} only include the $u_0>0$
solution, but there is an exact degeneracy with a model characterized by
$(u_0,\alpha)\mapsto-(u_0,\alpha)$, due to the symmetry of the lens. In practice, the other parameters
remain unchanged, so the physical properties of source and lens are identical.
Moreover, there is no close-wide $(s\mapsto 1/s)$ discrete degeneracy, for the
anomaly is due to the planetary caustic instead of the central caustic.
In other words, the source apparent trajectory passes in the middle of minor image caustics, in a region where
the magnification is lower than what would be observed if the lens were single. This feature is not 
easily reproductible with another lens geometry, which is, in part, the reason why there are not many competing models
for this event. \Fig{fig:caus1} shows the magnification residuals between the PSBL and PSPL
models, as well as magnification maps around the caustics computed with an adaptation of the library \texttt{luckylensing}\footnote{Published at \url{https://github.com/smarnach/luckylensing}.} \citep{Liebig2015}. The de-magnification regions appear in blue in this figure.

Although \eventname{} is a low-magnification event, an anomaly is clearly
identified at $t\approx6890$. Due to the possibility that this anomaly is
shaped by the effect of the physical size of the source, we introduce one more
model parameter: the source radius crossing time,
$\tS=\rho\,\tE=\thS/\murel$, where $\rho$ is the source angular radius
in units of $\thE$, \ie,
\begin{equation}\label{eq:rho}
  \rho=\frac{\thS}{\thE}\v
\end{equation}
with $\thS$ the source angular radius. Hereafter, we refer to the resulting
`finite-source binary-lens' model as `FSBL.' Finite source effects in
microlensing light curves are usually sensitive to the stellar limb darkening
\citep{Albrow.1999}, however only if the source star crosses the caustic, which is not
the case in \eventname.

We tried to extract constraints on $\tS$ in two ways. One using an \gls*{mcmc}
algorithm with no constraint on the parameters, and a large proposal step
function. The other fixing $\tS$ on a grid (25 nodes for
$0.04\,\days\leq\tS\leq1.05\,\days$), and searching for solutions with an
\gls*{mcmc} algorithm. These two approaches do not provide any useful limit on $\tS$.
In fact, the upper limit on $\tS$ provided by the light curve is found between $1.0$ and
$1.5\,\days$, corresponding to a $\chi^2$ increase of respectively $\sim 1$ and
$\sim 7$. In \Fig{fig:source}, this upper limit falls at the edge between the $3$ and 4-$\sigma$ confidence regions of the posterior distribution, \ie, the final constraint on $\tS$ exclusively comes from the galactic 
prior, rather than from the observations.
This result is mainly due to the source trajectory relative to lens. As shown in
\Fig{fig:caus1}, the PSBL solutions correspond to a caustic consisting in three
very small parts of the source plane (a `central caustic' and two `planetary
caustics'). Along its trajectory, the source remains almost equidistant from
the two planetary caustics, leaving the anomaly poorly magnified. 
However, a detailed analysis
unambiguously rules out the PSPL model by
$\Delta\chi^2=\chi^2_\mathrm{PSPL}-\chi^2_\mathrm{PSBL}=968$.

Despite the relatively short timescale of the event,
we also considered PSBL models, including the microlens parallax (hereafter
``$\text{PSBL}\oplus\piE$''). Although better by $\Delta\chi^2\approx -12$,
this model converges towards the unphysical large value
$\piE=\vpm{2.5}{0.9}{1.0}$ ($\piEN=\vpm{2.4}{0.9}{1.2}$ and $\piEE=0.3\pm0.1$),
and leaves the other parameters almost unchanged (all the parameters are within
1-$\sigma$ of the static model). This means that a model with parallax does not
change the interpretation of the lens.  The best $\text{PSBL}\oplus\piE$ model
is shown in green in \Fig{fig:lc1}. To assess whether the parallax detection is reliable,
we compute the Bayesian information criterion (BIC) to take into account the
effect of the additional free parameters in the models. The best PSBL model
with parallax is now marginally preferred by $\Delta\mathrm{BIC}\approx-0.03$.
As a consequence, we cannot claim that the microlens parallax can be reliably
measured using MOA observations of this event.

\section{Source and lens physical properties}\label{sec:properties}
\subsection{Nature of the source star}

As shown previously in \Sec{sec:models}, the source angular size is not detected in the light curve. Moreover, we do not have any color information about the source. Despite a lack of observational information, this section shows that the nature of the source can be determined: it is most likely a \gls*{rcg} star located in the Galactic bulge.

First, we build a \gls*{cmd} using the MOA-II $R$- and $V$-passband with stars within 2~arcmin around \eventname. Since the source brightness in $R$ band found in \Sec{sec:models} turns out to be the same as \gls*{rcg} stars, we assume that the source belongs to the \gls*{rcg}. Doing so, we implicitly reject the scenario with a foreground main sequence source. Although not impossible, this scenario is unlikely because the probability for a star to be lensed is proportional to $\DS^2$ \citep{Paczynski.1996}. Also, the foreground is much less populated by main sequence stars at a magnitude $I\approx16$, than the background.

\begin{figure}[tp]
  \begin{center}
    \includegraphics[scale=1.3]{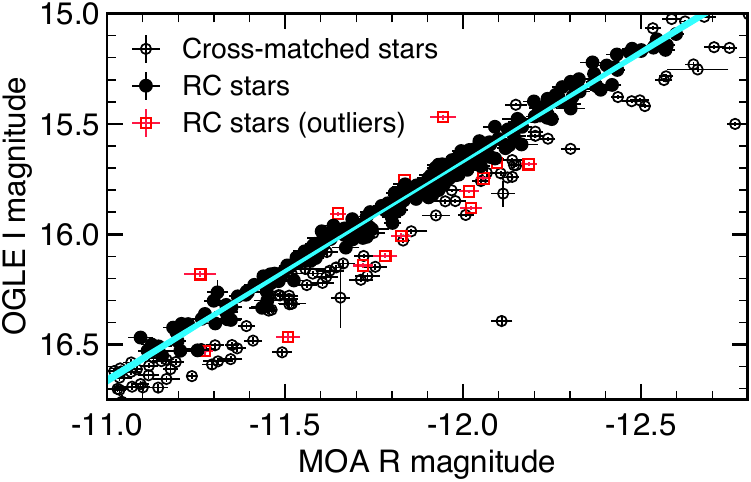}
    \caption{Empirical law between OGLE $I$ magnitude and the instrumental $\RMOA$ magnitude of stars located at an angular separation less than 2~arcmin from \eventname. \acrshort*{rcg} stars (black dots) only are used in the linear fit (except the outliers displayed as red squares). Black circles corresponds to stars that do not belong to the \acrshort*{rcg} (most of them are foreground main sequence stars). The cyan region is an envelop holding 100,000 randomly chosen samples.}\label{fig:calib}
  \end{center}
\end{figure}

The second step is to calibrate the instrumental MOA-II $\RMOA$ magnitude by cross-referencing stars from the MOA-II \texttt{DOPHOT} catalog with stars in the OGLE-III catalog. We use these stars to build a catalog with magnitudes in the standard Kron–Cousins $I$ and Johnson $V$ passbands \citep{Szymanski.2011}. 
In OGLE-III catalog, we identify 7446 stars located less than 2~arcmin from \eventname, while we find 1222 stars in the MOA-II catalog. \Fig{fig:cmd} shows the resulting OGLE-III $(V-I, I)$ \gls*{cmd}. Following the method described in \citet{Nataf.2016}, we identify \gls*{rcg} stars to derive their centroid (red circle in \Fig{fig:cmd}).
A total of 818 stars are cross-matched, including 251 \gls*{rcg} stars (see \Fig{fig:cmd}). Since we assume a source that belongs to the \gls*{rcg}, we select those stars to derive an empirical linear law between OGLE-III $I$ and MOA-II $\RMOA$ magnitudes. \Fig{fig:calib} displays the aforementioned cross-matched stars, the \gls*{rcg} stars used in the linear fit, and outliers. The outliers are identified by following the methodology described in Section 3 of \citet{Hogg.2010}, taking alternatively into account the error bars of $I$ and $\RMOA$. We remove from the final fit the \gls*{rcg} stars that are classified as outliers in both cases. During this process, we note that an underestimate of the photometric error bars seem to be responsible for being classified as an outlier. The final linear fit is then performed following Section~7 of \citet{Hogg.2010}, taking into account two-dimensional uncertainties. The resulting empirical law reads, $I = a_0 + a_1 \RMOA$,
with $a_0 = 27.58\pm 0.06$ and $a_1 = 0.992\pm 0.005$. These values correspond to the median of the marginalized posterior distributions (\ie, the two values do not necessarily represent a good fit), sampled with a \gls*{mcmc} algorithm. The error bars display the 68.3 per cent credible interval around the median, derived from the 16 and 84 per cent percentile of the corresponding marginalized posterior distribution.
\Fig{fig:calib} shows the envelop that holds 100,000 randomly chosen samples.

The third step is to use the calibration law found in step 2 to derive the $I$-band source magnitude. \Fig{fig:cmd} shows the source location in the OGLE \gls*{cmd} (cyan point), when its color is assumed to be the same as the \gls*{cmd} centroid at the corresponding brightness $\pm 0.4$~mag, and with the same dispersion. In practice, for each value of the source brightness derived from the previous step, the source color is described by a Gaussian distribution, which mean coincides with the centroid of \gls*{rcg} stars, and with a standard deviation derived from the color dispersion of \gls*{rcg} stars that have the same brightness as the source $\pm 0.4$~mag. Under this assumption, the following paragraphs explain how we estimate the source radius from its brightness and color.

To do so, we measure the extinction and reddening of stars within 2 arcmin around \eventname. The centroid of the \gls*{rcg} stars is $\vmirc = 2.27\pm0.02$ and $\Irc=15.8\pm0.1$. The absolute magnitude of a source located in the Galactic bulge is $M_\mathrm{I,RCG} = -0.17\pm0.05$ \citep{Chatzopoulos.2015, Nataf.2016} and its intrinsic color is $(V-I)_\mathrm{RCG,0}=1.06$ \citep{Bensby.2013}. We use a new Galactic model \citep{Koshimoto2021,Koshimoto2021code} to estimate the distance modulus of the source, $\mu = \vpm{14.60}{0.21}{0.15}$, corresponding to $\DS = \vpm{8.34}{0.86}{0.57}\,\kpc$. As expected, these values are consistent with the assumption we made of a \gls*{rcg} source.
The new Galactic model improves several aspects of previous ones \citep{Bennett.2014,Zhu.2017.gal}, for instance by taking into account the change in the velocity dispersion within the disk, with respect to the distance to the Galactic center.
Since the extinction and reddening mostly occurs during the first kiloparsecs away from Earth, the dereddened source magnitude is $I_\mathrm{s,0} = I_s + M_\mathrm{I,RCG} + \mu - \Irc$, \ie, $I_\mathrm{s,0} = 14.4\pm0.3$, and $\vmi{s,0}=1.06\pm0.14$. The corresponding extinction $A_I=1.39^{+0.16}_{-0.22}$, and color excess $E(V-I)=1.2\pm0.1$ found are in good agreement with the $A_I=1.47$ and $E(V-I)=1.21$ derived from \cite{Gonzalez.2012}.

\begin{figure}[tp]
  \begin{center}
    \includegraphics[scale=1.3]{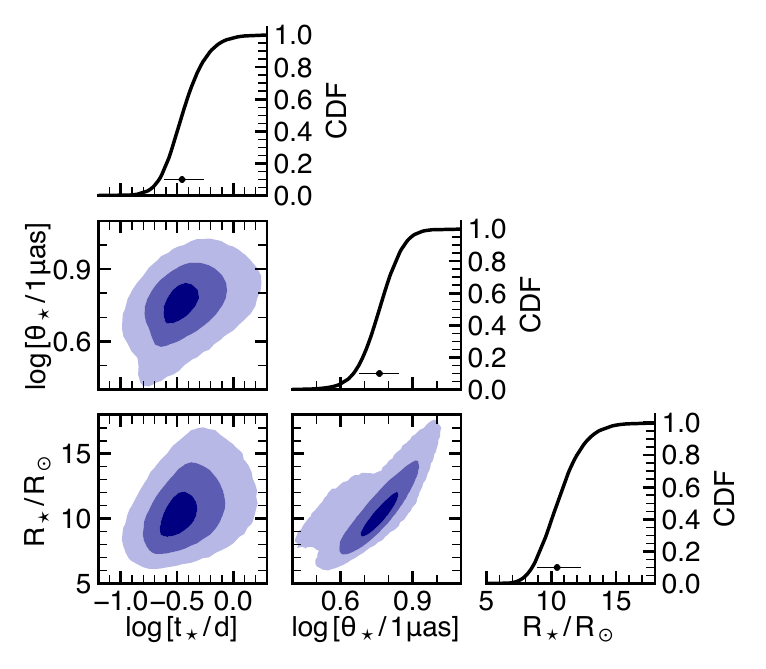}
    \caption{Correlation between the parameters for the best-fit model. The three shaded areas show the 1-3$\sigma$ confidence regions, respectively, from the darkest to lightest color. Plots in the diagonal displays the marginal cumulative distribution function of each parameter (solid line), the median of the distribution (dot), and the 68.3 per cent credible interval centered on the median. Plot prepared using the python package \texttt{MOAna} \citep{Ranc2020moana}.}\label{fig:source}
  \end{center}
\end{figure}

\begin{table}[tp]
{\centering
    \caption{Lens and Source Properties Derived from the Bayesian Analysis of \Sec{sec:properties}}\label{tab:lens}
{\footnotesize
\begin{tabular}{@{} l c c @{}}\toprule
    Parameter                   & MCMC results$^{\dagger}$   & Units\\ \midrule
Host mass $M_1$                & $\vpm{0.31}{0.36}{0.19}$ & $\Msun$ \\[2pt]
Planet mass $M_2$              & $\vpm{1.9}{2.2}{1.2}$  & $\Mjup$ \\[2pt] 
Projected separation $a_\perp$ & $0.75\pm0.24$ & $\au$   \\[2pt]
Deprojected separation $a$     & $0.96\pm0.31$ & $\au$   \\[2pt]
Lens distance $\DL$            & $\vpm{7.2}{0.6}{1.7}$  & $\kpc$  \\\midrule
Einstein radius$^{\ddagger}$ $\thE$           & $\vpm{0.24}{0.09}{0.08}$ & $\mas$           \\[2pt]
Lens-source proper motion$^{\ddagger}$ $\murelg$ & $\vpm{6.0}{2.3}{2.0}$ & $\mas\,\yr^{-1}$ \\[2pt]
Source magnitude    $\Iso$                    & $14.4\pm0.3$                   & mag              \\[2pt]
Source color  $\vmi{s,0}$                     & $1.06\pm0.14$                  & mag              \\[2pt]
Extinction $A_I$                              & $\vpm{1.39}{0.16}{0.22}$       & mag              \\[2pt]
Reddening $E(V-I)$                            & $1.2\pm0.1$                    & mag              \\[2pt]
Source angular radius $\thS$                  & $\vpm{5.7}{1.2}{1.0}$          & $\uas$           \\\bottomrule
\end{tabular}}\par}
{\footnotesize%
    $^{\dagger}$ Median of the marginalized posterior distributions, with error bars displaying the 68 per cent credible interval around the median.\\
    $^{\ddagger}$ Galactic prior.
}
\end{table}

Finally, the angular source size can be estimated using the empirical relation \citep{Boyajian.2014},
\begin{equation}\label{eq:empirical}
    \log\left(\frac{2\thS}{\mas}\right) = 0.501414 - 0.2I_\mathrm{s,0} + 0.419685\vmi{s,0}\v 
\end{equation}
inferred from stars with colors corresponding to $3900\,\mathrm{K} < T_\mathrm{eff} < 7000\,\mathrm{K}$ \citep{Bennett.2017}. In \Eq{eq:empirical}, `$\mas$' denotes milli-arcsec. The resulting source angular radius yields the source radius, $R_\star=\thS\DS$, and the source radius crossing time, $\tS=\thS/\murel$, shown in \Fig{fig:source}. With $\thS=5.8^{+1.2}_{-1.0}\,\uas$ (`$\uas$' denotes micro-arcsec) and $R_\star=10.5^{+1.8}_{-1.5}\,\Rsun$, we check that the source is a red giant star of the Galactic bulge, as we assumed.

The exact origin of the blend flux remains unknown. The ratio of the blend flux to the source flux for the binary-lens models, $\fbl{R}/\fsl{R}$, is $0.4\pm0.2$ (see \Tab{tab:model_parameters}). It may be due to one or several stars, including the lens, blended into the point spread function. As a consequence, the blend flux cannot be used to characterize further the nature of the lens.

\subsection{Nature of the lens}

The main difficulty of the lens characterization is that the light-curve modeling returns
only one parameter that is sensitive to the mass and distance, namely, the Einstein timescale
defined in \Eq{eq:tE}. The mass-distance dependence of $\tE$ appears in the expression of
the angular Einstein radius; \ie
\begin{equation}\label{eq:thE}
    \qquad \thE = \sqrt{\frac{4G\ML}{c^2\DS}\left(\frac{\DS}{\DL}-1\right)}\v
\end{equation}
where $M$ is the lens mass, $\DS$ and $\DL$ are the distances to the source and lens, $c$ is the
speed of light, and $G$ is the gravitational constant.
We use a galactic model of the Milky Way to predict the distribution of angular Einstein radii, source distances and lens-source relative parallaxes as introduced in \Eq{eq:piE},
\begin{equation}\label{eq:pirel}
    \pirel = \frac{1\,\au}{\DL} - \frac{1\,\au}{\DS}\v
\end{equation}
from the event coordinates. This model assumes that all stars have an equal planet hosting probability. Then, we use these predictions as priors to derive the total mass of the lens using
Equations~\ref{eq:thE} and~\ref{eq:pirel}, \ie,
\begin{equation}\label{eq:mass}
    M = 0.1228\,\Msun\left(\frac{\thE}{1\,\mas}\right)^2\left(\frac{\pirel}{1\,\mas}\right)^{-1}\v
\end{equation}
and the distance to the lens,
\begin{equation}
\DL = 1\,\kpc \left(\frac{\pirel}{1\,\mas} + \left(\frac{\DS}{1\,\kpc}\right)^{-1}\right)^{-1}\p
\end{equation}
Since the angular Einstein radius measurements via microlensing is typically $> 1$ per cent, the precision of the lens mass estimation is expected to be $> 2$ per cent; we choose the significant digits of the constant in Equation~\ref{eq:mass} accordingly.
Finally, the host-star and planet masses can be found from the measurement of the mass ratio in \Sec{sec:binary-models}, \ie,
\begin{equation}
    M_1=\frac{M}{1+q} \qquad \text{and} \qquad M_2=\frac{qM}{1+q}\p
\end{equation}
The results of the Bayesian analysis are summarized in \Tab{tab:lens}.

The lens likely consists of a $\vpm{1.9}{2.2}{1.2}\,\Mjup$ Jupiter-mass planet orbiting a $\vpm{0.31}{0.36}{0.19}\,\Msun$ M-dwarf star. As expected, the lack of source size measurement is responsible for large uncertainties on
the mass of each component of the lens system. The planet-host star projected separation is $0.75\pm0.24\,\au$. If we assume a circular orbit, this value translates into a mean semi-major axis $0.96\pm0.31\,\au$. This planetary system lies at a distance $\vpm{7.2}{0.6}{1.7}\,\kpc$.

\begin{figure}[tp]
  \begin{center}
    \includegraphics[scale=1.6]{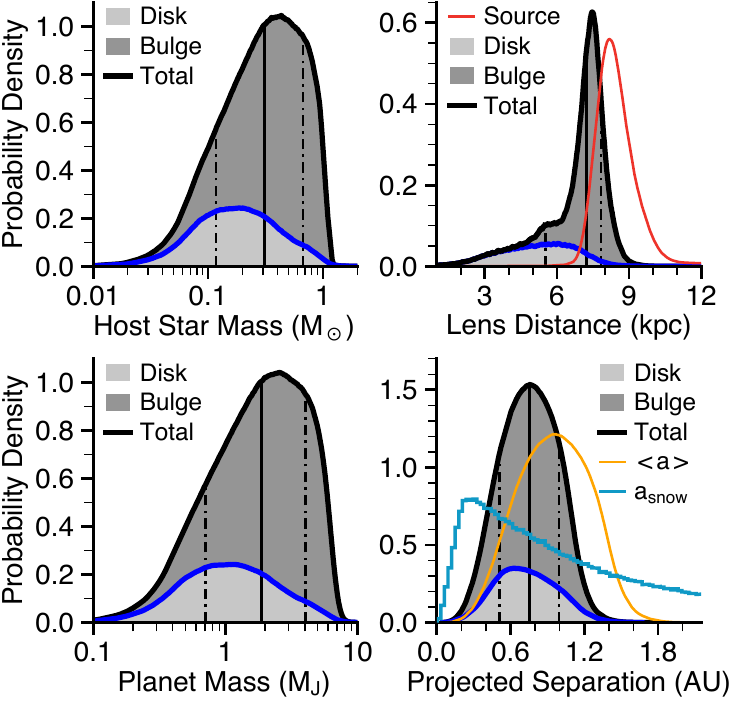}
    \caption{One-dimensional marginalized posterior probability density function of the host star mass (upper left panel), distance to the lens (upper right panel), planet mass (lower left panel), and projected separation (lower right panel). Two shaded areas are separated by a dark blue line. They show the contribution of the thin and thick disks (light gray), and the spheroid and bulge (dark gray) to the posterior distribution (black line). The upper right panel also displays the prior distribution on the source distance (red dashed line), derived from the galactic model. The lower right panel includes the probability density function of the deprojected separation (orange line), $<a>$, and snow line position (light blue line), $\asnow$.}\label{fig:lens}
  \end{center}
\end{figure}

In \Fig{fig:lens}, the light gray shading indicates the thin and thick disk contribution to the posterior distribution (black solid curve), while the dark gray shading indicated the spheroid and bulge contribution to the posterior distribution. Although these profiles raise the possibility of a lens lying in the disk, they suggest that a bulge lens is much more likely.

\section{Summary and Discussion}\label{sec:conclusion}

We have reported the discovery of a new Jupiter-mass planet,
\eventnameLb, discovered through a low magnification anomaly during
the microlensing event \eventname. The anomaly was due to the source
star passing in between the two off axis components of a close caustic,
consistent with a planet-to-host-star mass ratio
$q=5.75_{-0.42}^{+0.45}\times10^{-3}$. Since microlensing in the Milky
Way is most often caused by M-dwarfs lenses, this mass ratio
corresponds typically to the domain of giant planets. The projected
separation between the planet and the host star is $s=0.47\pm0.02$
Einstein radius. The $s\leftrightarrow1/s$ degeneracy does not exist
for this event, because the anomaly is not due to the central caustic.
An exact geometrical degeneracy exists, leaving the lens physical
parameters unchanged, though.

Due to its low magnification (maximum $A\approx1.4$), and anomaly
occurring at an extremely low magnification ($A\approx1.06$), we did
not detect features resulting from the angular size of the source.
Without this measurement, we cannot use the light curve to measure
$\thE$. However, we used a Galactic model to predict the distribution of
the Einstein radius, source distance, relative lens-source proper motion,
and microlens parallax. The resulting constraints on the lens physical properties
are weak, but a low mass ratio in conjunction with a likely low-mass host enables 
us to put the mass of the companion in the planetary mass regime.

Including planets like MOA-2014-BLG-472Lb in statistical studies on planet demography is crucial for the completeness of planetary occurrence rates. The event \eventname{} (including the anomaly) was intensively monitored by the MOA survey. Interestingly, although the physical parameters of MOA-2014-BLG-472Lb are not tightly constrained, the mass ratio, $q$, and the projected separation, $s$, are both precisely measured, and not degenerate. Events without close-wide degeneracy are not so common in statistical analyses. Since \eventname{} does not suffer from it, it is an important add on to the new MOA sample of microlensing planets, that will be used in the next statistical analysis.

\section*{Data Availability}

The original data underlying this article can be accessed from the NASA Exoplanet Archive MOA Resources, \url{https://exoplanetarchive.ipac.caltech.edu/docs/MOAMission.html}. The derived data generated in this research will be shared on reasonable request to the corresponding author until they are added to the NASA Exoplanet Archive, at \url{https://exoplanetarchive.ipac.caltech.edu/docs/contributed_data.html}.

\acknowledgments

The authors thank the anonymous referee, whose comments and suggestions improved the manuscript. CR thanks Anne Shrestha for stimulating discussions.
This work was carried out within the framework of the ANR project COLD-WORLDS supported by the \emph{Agence Nationale de la Recherche} (French National Agency for Research) with the reference ANR-18-CE31-0002.
The work of CR, DPB, NK, YH and AB was supported by NASA under award number 80GSFC17M0002.
JS acknowledges support from the National Science Center, Poland grant MAESTRO 2014/14/A/ST9/00121.
The MOA project is supported by JSPS KAKENHI grant No. 19KK0082 and 20H04754.
This research has made use of the NASA Exoplanet Archive, which is operated by the California Institute of Technology, under contract with the National Aeronautics and Space Administration under the Exoplanet Exploration Program.

\software{Astropy \citep{Astropy.2018}, Luckylensing (\url{https://github.com/smarnach/luckylensing}), Matplotlib \citep{matplotlib}, MOAna \citep{Ranc2020moana}, NumPy \citep{Oliphant2015}, SciPy \citep{scipy}.}

\bibliography{references}

\end{document}